\documentclass[12pt]{article}
\usepackage{graphicx}%
\usepackage{amsmath}
\usepackage{amsfonts}%
\usepackage{amssymb}
\begin{document}

\begin{center}
{\bfseries Charged pion polarizability in the nonlocal quark model
of Nambu--Jona-Lasinio type}

\vskip 5mm

A.~E.~Radzhabov$^{\dag}$ and M.~K.~Volkov$^{\ddag}$

\vskip 5mm

{\small
{\it
Bogoliubov Laboratory of Theoretical Physics, \\
Joint Institute for Nuclear Research, 141980 Dubna, Russia
}
\\
$\dag$ {\it
E-mail: aradzh@thsun1.jinr.ru
}
\\
$\ddag$ {\it
E-mail: volkov@thsun1.jinr.ru
}}
\end{center}

\vskip 5mm

\begin{center}
\begin{minipage}{150mm}
\centerline{\bf Abstract} The polarizability of a charged pion
is estimated in the framework of the nonlocal chiral quark model
of the Nambu--Jona-Lasinio type. Nonlocality is described by
quark form factors of the Gaussian type. It is shown that the
polarizability in this model is very sensitive to the form of
nonlocality and choice of the model parameters.
\end{minipage}
\end{center}

\vskip 10mm

Recently, the interest to direct determination of the pion
polarizability in Primakoff scattering has been renewed. Now the
new experiment at CERN is prepared by the collaboration COMPASS
where it is expected to obtain the statistics factor 6000 higher
than in the previous experiment performed by the IHEP and JINR
group in Protvino \cite{exp1} at the beginning of the 80s.
However, the technical possibilities of this experiment allow
one to obtain only very rough estimations of the pion
polarizability with large error bars
\begin{align}
\alpha_{\pi}=-\beta_{\pi}=8.54 \pm 1.76 \pm 1.51,\label{ex1}
\end{align}
where $\alpha_{\pi}$ and $\beta_{\pi}$ are electric and
magnetic polarizabilities of the charged pion. Hereafter we
express the polarizability in the units of $10^{-42}$cm$^3$.
The new experiment gives an opportunity to obtain a value of
the pion polarizability with good accuracy
\cite{Moinester:2003rb}.

What concerns the theoretical aspect of this problem for the
past years, there are a lot of estimations of this quantity
made by many authors in different theoretical models,
in particular, one of the authors of this article (MKV)
together with V.~N.~Pervushin obtained estimations of the pion
polarizability in the framework of the nonlinear chiral
model in 1975 \cite{Pervushin:xq}. After that, analogous
calculations were performed by him in the quark linear
sigma model of Nambu--Jona-Lasinio(NJL) type
\cite{Volkov:nk,Volkov:zb}. The estimation obtained in the
above-mentioned works corresponds to the experimental result
(\ref{ex1}).

In this short note, we want to return to this question and
estimate the pion polarizability in the nonlocal
model of the NJL type \cite{RV}, where nonlocality
described by quark form factors of the Gaussian type.

\begin{figure}
\begin{center}
\resizebox{0.40\textwidth}{!}{\includegraphics{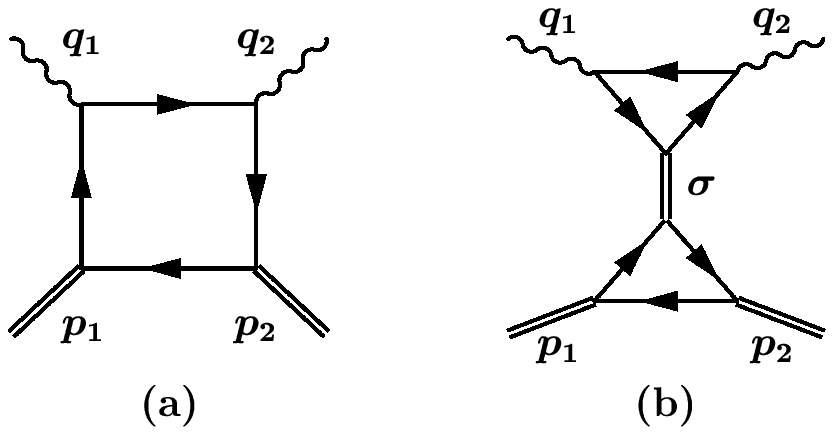}}
\caption{Diagrams describing charge pion polarizability: a)
box diagrams; b) $\sigma$ pole diagrams.}
\label{polfig}
\end{center}
\end{figure}

As it is shown in \cite{Volkov:zb}, in the local NJL model
the main contribution to the pion polarizability in the leading
order of $1/N_c$ expansion stems from two types of
diagrams: the diagrams with light intermediate $\sigma$-meson,
and the box diagrams (see fig.\ref{polfig}).
The contribution from the diagrams with other resonances
(heavier scalars, vectors and axial-vectors) is smaller
than $3\%$ and can be neglected in the present consideration.

The contribution of the box diagrams to the structure part
of the Compton effect takes the form
\begin{align}
A^{\mu\nu}_{(a)}=&\frac{\alpha}{9 \pi f_\pi^2}
(g^{\mu\nu}(q_1 \cdot q_2) - q_2^\mu q_1^\nu)\nonumber,
\end{align}
where $q_1,q_2$ are the momenta of the incoming(outgoing)
photons, $\alpha=1/137$ is the fine structure constant,
$f_\pi=93$ MeV is the weak pion decay constant. The contribution
of the $\sigma$ pole diagram can be factorized into three parts
\begin{align}
A^{\mu\nu}_{(b)}=&A_{\sigma \pi \pi} D_\sigma A_{\sigma
\gamma \gamma}^{\mu\nu},\nonumber
\end{align}
where the $\sigma \pi \pi$ vertex is $A_{\sigma \pi \pi}=4 m g Z
={4 m^2 Z^{1/2}}/{f_\pi}$, the $\sigma$ meson propagator is
$D_\sigma \approx 1/M_\sigma^2$ and the $\sigma \gamma \gamma$
vertex is \cite{Dorokhov:rv}
\begin{align}
A^{\mu\nu}_{\sigma \gamma \gamma}=(g^{\mu\nu}(q_1 \cdot q_2) -
q_2^\mu q_1^\nu) \frac {10 \alpha Z^{-1/2}}  {9 \pi f_\pi}
\,,\nonumber
\end{align}
here $m=280$ MeV is the constituent quark mass after taking into
account the $\pi-a_1$ mixing \cite{Volkov:zb}, $g=m Z^{-1/2}
/f_\pi$ is the $\sigma$ meson coupling constant, factor $Z$ is
$Z=({1-6m^2/M_{a_1}^2})^{-1} \approx 1.4$ ($M_{a_1}=1.26$ GeV is
the mass of the $a_1$ meson), $M_\sigma$ is the $\sigma$ meson
mass $M_\sigma=\sqrt{4m^2+M_\pi^2}\approx 2m$ and $M_\pi$ is the
pion mass. As a result, the pion polarizability takes the form
\begin{align}
\alpha_\pi=-\beta_\pi=\frac{5\alpha}{9 \pi M_\pi
f_\pi^2}\left(1-\frac{1}{10}\right)=7.3 \label{polNJL},
\end{align}
One can see that the polarizability does not depend on the model
parameters.

Let us emphasize that the contribution of the box diagrams makes
up only 10 \% of the contribution of the $\sigma$ pole
diagrams\footnote{It is worth noting that a similar situation
with relative contributions from the $\sigma$ pole and box
diagrams takes place in nonlocal model (see, e.g.,
\cite{Ivanov:1991kw}).}.

Our calculation shows that in the nonlocal model the choice of
the form-factor $f(p)$ and the model parameters have more
influence on the pion polarizability than taking into account
the box diagrams. Therefore, in this note we restrict ourselves
only with the estimations of the $\sigma$ pole diagrams
contribution.

\begin{figure}
\begin{center}
\resizebox{0.80\textwidth}{!}{\includegraphics{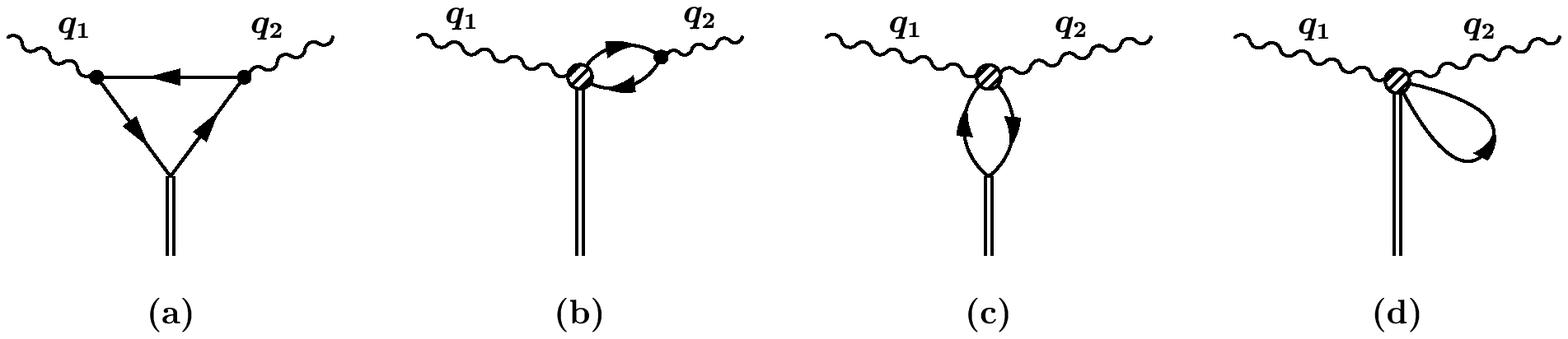}}
\caption{Diagrams describing vertex $\sigma \gamma \gamma$.}%
\label{sggfig}%
\end{center}
\end{figure}

In the nonlocal, the model quark mass depends on momentum. In
\cite{RV}, the following representation for the quark propagator
is proposed
\begin{eqnarray}
\frac{m^2(p)}{m^2(p)+p^2} = \exp
\left(-{p^2}/{\Lambda^2}\right),\label{neqq1}
\end{eqnarray}
where $\Lambda$ is the parameter of nonlocality. As a result,
the pole part of the quark propagator has no singularities on
the whole real axis, what leads to the quark confinement. The
quark mass function takes the form \cite{RV}:
\begin{equation}
m^2(p)=\left(\frac{p^{2}}{\exp\left(
{p^{2}}/{\Lambda^{2}}\right)-1}\right).\label{M(p)}
\end{equation}
From this equation one can see that $m(0)=\Lambda$. Then, from
the condition that the weak pion decay constant equals $93$ MeV,
it follows that $m(0)=\Lambda=340$ MeV. For the $\sigma \pi \pi$
vertex and the $\sigma$ meson mass we have $A_{\sigma \pi
\pi}=1.57$ GeV, $M_\sigma=420$ MeV \cite{RV}. The vertex $\sigma
\gamma \gamma$ is more complicated. We can expressed it as
\begin{align}
A_{\sigma \gamma \gamma}^{\mu\nu}&=\alpha C_{\sigma \gamma
\gamma}(g^{\mu\nu}(q_1 \cdot q_2) - q_2^\mu q_1^\nu).
\nonumber
\end{align}
Due to the $\mathcal{P}$exp factor in the action
\cite{RV,DoLT98}, there are additional nonlocal photon vertices.
The reason for appearing of these vertices is the momentum
dependence of the quark mass and the meson-quark vertices. The
technique of obtaining of these vertices can be found in
\cite{DoLT98,Terning:1991yt}. As a result, there are four types
of diagrams describing the process $\sigma \gamma \gamma$,
fig.\ref{sggfig}, only sum of them is gauge invariant.
As a result, we have\footnote{It is easy to see that $A_{\sigma
\pi \pi}$, $C_{\sigma\gamma\gamma}$ noticeably decrease in
comparison with the local NJL model(see also table \ref{tab1} ).
We would like to emphasize that similar situation takes place in
the other nonlocal model with form factors of the Gaussian type
\cite{Faessler:2003yf}.}
\begin{align}
C_{\sigma\gamma\gamma}&=0.86\,\,\mathrm{GeV^{-1}},\nonumber
\end{align}
and pion polarizability is
\begin{align}
\alpha_\pi=1.5\nonumber.
\end{align}

Let us notice that the calculations of the pion polarizability
in a similar nonlocal model of the NJL type with quark
form-factors of the Gaussian type is performed in \cite{DB} by
using the chiral sum rule method. Those form-factors lead to the
following momenta dependence of quark mass function
\begin{equation}
m(p)=m_0 \exp\left(-2{p^{2}}/{\Lambda^{2}}\right).\nonumber
\end{equation}
In contrast to the previous model, here the only one condition
is used $f_\pi=93$ MeV for fixing two main model parameters
$m_0$, $\Lambda$. This keeps some freedom in choosing model
parameters. Particularly, in \cite{DB} the model parameters
$m_0=300$ MeV, $\Lambda=1.085$ GeV are used. As a result, the
values of the pion polarizability obtained by the sum rule
method is $\alpha_\pi=3.6$. On the other hand, we can carry out
calculations of the pion polarizability using the $\sigma$ pole
diagrams. We have
\begin{align}
A_{\sigma \pi \pi}=&1.47\,\,\mathrm{GeV},\\
M_\sigma=&398\,\,\mathrm{MeV},\nonumber\\
C_{\sigma\gamma\gamma}=&1.75\,\,\mathrm{GeV^{-1}},\nonumber
\end{align}
and, as a result, we obtain $\alpha_\pi=3.3$. It worth noting
that if $m_0$ is equal to $340$ MeV, as in model \cite{RV} (see
eq.(\ref{M(p)})), the model quantities equal
\begin{align}
\Lambda=&0.92\,\,\mathrm{GeV}\nonumber\\
A_{\sigma \pi \pi}=&1.57\,\,\mathrm{GeV},\\
M_\sigma=&421\,\,\mathrm{MeV},\nonumber\\
C_{\sigma\gamma\gamma}=&1.16\,\,\mathrm{GeV^{-1}},\nonumber
\end{align}
the polarizability of the pion is $\alpha_\pi=2.0$ (sum rule
estimation is $\alpha_\pi=2.7$). If one takes $m_0=280$ MeV, as
in the local NJL model, one can obtain\footnote{Note that in the
local NJL model, $\Lambda$ equals $1.25$ GeV \cite{Volkov:zb}. }
\begin{align}
\Lambda=&1.187\,\,\mathrm{GeV},\nonumber\\
A_{\sigma \pi \pi}=&1.39\,\,\mathrm{GeV},\\
M_\sigma=&384\,\,\mathrm{MeV},\nonumber\\
C_{\sigma\gamma\gamma}=&2.1\,\,\mathrm{GeV^{-1}},\nonumber\\
\alpha_\pi=&3.9\,(\alpha_\pi=5.0\,\mathrm{from\,the\,sum\,rules}).\nonumber
\end{align}
As we can see, in the nonlocal model there is a strong
dependence of the pion polarizability on the form of
nonlocality. We summarize theoretical results in the table
\ref{tab1}.

\begin{table}
\begin{center}
\begin{tabular}[]{|c|c|c|c|c|c|c|c|}
\hline
model&$m(0)$&$\Lambda$&$M_\sigma$&$A_{\sigma \pi \pi}$&$C_{\sigma\gamma\gamma}$&$\alpha_{\pi}$ sigma pole&$\alpha_{\pi}$ sum rules\\
&MeV&MeV&MeV&GeV&GeV$^{-1}$&$10^{-42}\,\mathrm{cm}^{3}$&$10^{-42}\,\mathrm{cm}^{3}$\\
\hline
\cite{RV}&$340$&$340$&$420$&$1.57$&$0.86$&$1.5$&$2.0$ \\
\hline
\cite{DB}&$340$&$920$&$421$&$1.57$&$1.16$&$2.0$&$2.7$ \\
\hline
\cite{DB}&$300$&$1085$&$398$&$1.47$&$1.75$&$3.3$&$3.6$ \\
\hline
\cite{DB}&$280$&$1187$&$384$&$1.39$&$2.1$&$3.9$&$5.0$ \\
\hline
\cite{Volkov:zb}&$280$&$1250$&$577$&$4$&$3.2$&$8.1$& \\
\hline
\end{tabular}
\end{center}
\caption{Theoretical values of pion polarizability.}
\label{tab1}
\end{table}

The experimental values of pion polarizability are given in the
table \ref{tab2}.
\begin{table}
\begin{center}
\begin{tabular}[]{|c|c|}\hline
$\alpha_\pi^{\mathrm{exp}}$ &  experiment\\
\hline
$8.54\pm 1.76 \pm 1.51$&\cite{exp1} \\
\hline
$3.3\pm 0.6$&\cite{exp3} \\
\hline
$3.31\pm 0.45$&\cite{exp4} \\
\hline
$3.4 \pm 1.11$&\cite{exp5} \\
\hline
\end{tabular}
\end{center}
\caption{Experimental values of the pion polarizability:
\cite{exp1} measurement via Primakoff scattering; \cite{exp3}
deduced from the processes $\gamma\gamma\to\pi\pi$; \cite{exp4},
\cite{exp5}  obtained from the sum rules for vector and
axial-vector correlation functions in $\tau$ decays.}
\label{tab2}
\end{table}

To conclude, we consider the estimation of the pion
polarizability in the framework of the local and nonlocal models
of the NJL type. It is shown that the pion polarizability in the
nonlocal model is noticeably smaller than in the local NJL
model, and is very sensitive to the form of nonlocality and
choice of the model parameters. Planned experiments must give
more accurate values of the pion polarizability. We hope that
these data allow us to chose more realistic version of the
nonlocal model.

The authors thank I. O. Cherednikov, A. E. Dorokhov, S. B.
Gerasimov, M. A. Ivanov and E. A. Kuraev for useful discussions.
The work is partially supported by RFBR Grant no. 02-02-16194
and the Heisenberg--Landau program.

\end{document}